\documentclass[12pt]{article}
\usepackage{graphics}
\usepackage{amsfonts}
\usepackage{amsmath}
\usepackage{amssymb}

\def\\{\hfill\break} \let\==\equiv

\let\0=\noindent

\def\qed{\hfill\raise1pt\hbox{\vrule height5pt width5pt depth0pt}}

\def\be{\begin{equation}}
\def\ee{\end{equation}}
\def\bea{\begin{eqnarray}}\def\eea{\end{eqnarray}}

\begin{document}
\title{Making Sense of Bell's Theorem and Quantum Nonlocality\footnote{This is an expanded version of a talk given at the 2016 {\it Princeton-TAMU Symposium on Quantum Noise Effects in Thermodynamics, Biology and Information} \cite{Bou16}.}}
\author {Stephen Boughn{\small\it\thanks{sboughn@haverford.edu}}\author{Stephen Boughn}
\\[2mm]
~ \it Princeton University, Princeton NJ 08544 \\
~ \it Haverford College, Haverford PA 19041}

\date{{\small   \LaTeX-ed \today}}

\maketitle

\begin{abstract}
Bell's theorem has fascinated physicists and philosophers since his 1964 paper, which was written in response to the 1935 paper of Einstein, Podolsky, and Rosen. Bell's theorem and its many extensions have led to the claim that quantum mechanics and by inference nature herself are nonlocal in the sense that a measurement on a system by an observer at one location has an immediate effect on a distant {\it entangled} system (one with which the original system has previously interacted). Einstein was repulsed by such ``spooky action at a distance" and was led to question whether quantum mechanics could provide a complete description of physical reality. In this paper I argue that quantum mechanics does not require spooky action at a distance of any kind and yet it is entirely reasonable to question the assumption that quantum mechanics can provide a complete description of physical reality. The magic of entangled quantum states has little to do with entanglement and everything to do with superposition, a property of all quantum systems and a foundational tenet of quantum mechanics.

\\ {\small{Keywords: Quantum nonlocality $\cdot$ Bell's theorem $\cdot$ Foundations of quantum mechanics $\cdot$ Measurement problem}}

\end{abstract}

\maketitle

\section{Introduction}\label{INTRO}

In the 80 years since the seminal 1935 Einstein-Podolsky-Rosen paper (EPR)\cite{EPR35}, physicists and philosophers have mused about what Einstein referred to as ``spooky action at a distance". Bell's 1964 analysis of an EPR-type experiment, ``On the Einstein Podolsky Rosen Paradox"\cite{Bel64}, has been cited more than 10,000 times and most of these have been in the last decade. Many of these describe experimental work on testing quantum mechanical predictions related to entangled states and, as an experimentalist, I have great admiration for much of this work. On the other hand many, if not most, of the remainder are theoretical and philosophical papers trying to come to grips with what is often referred to as the nonlocality of quantum mechanics. According to David Mermin \cite{Mer85}, physicists' views of Bell's theorem fall ``between the majority position of near indifference and the minority position of wild extravagance." At the ``wild extravagance" extreme is, perhaps, Henry Stapp  who pronounced, ``Bell's theorem is the most profound discovery of science."\cite{Sta75}  Mermin advocates a balanced position between these two extremes. A distinguished Princeton physicist (Arthur Wightman?) concurred with Mermin's view but added, ``Anybody who's not bothered by Bell's theorem has to have rocks in his head."\cite{Mer85}. My view also lies in this middle ground but, I suspect, is closer to ``near indifference" than Mermin's.

Bell's theorem provides a general test, in the form of an inequality, that the results of an EPR type gedanken experiment must satisfy if it is describable by any classical theory, even one with hidden variables, so long as such a model is locally causal. Bell then proceeded to demonstrate how the predictions of quantum mechanics violate this inequality. His conclusion was that any hidden variable theory designed to reproduce the predictions of quantum mechanics must necessarily be nonlocal and allow superluminal interactions. In Bell's words\cite{Bel64},
\begin{quote}
In a theory in which parameters are added to quantum mechanics to determine the results of individual measurements, without changing the statistical predictions, there must be a mechanism whereby the setting of one measurement device can influence the reading of another instrument, however remote. Moreover, the signal involved must propagate instantaneously, so that such a theory could not be Lorentz invariant.
\end{quote}

While I suspect that most physicists never doubted the veracity of the predictions of quantum theory, 8 years later Freedman and Clauser\cite{Fre72} provided experimental verification of a violation of Bell's inequality. As a consequence, a large class of local hidden variable theories were immediately ruled out\footnote{David Bohm, building on de Broglie's notion of a pilot wave, had already created a nonlocal hidden variable theory\cite{Boh52} that reproduced all the predictions of non-relativistic quantum mechanics. Because of its nonlocality, it does not violate Bell's inequality.}. On the other hand, a demonstration of a violation of Bell's inequality certainly does not immediately imply that nonlocality is a characteristic feature of quantum mechanics let alone a fundamental property of nature. Nevertheless, many physicists and philosophers of science do harbor this belief. In fact, Bell later described the violation of his inequalities as pointing to the``gross nonlocality of nature"\cite{Bel75}.  Statements like ``If one assumes the world to be real, then Bell-experiments have proven that it is nonlocal"\cite{Wis06} and ``Bell's theorem asserts that if certain predictions of quantum theory are correct then our world is nonlocal...experiments thus establish that our world is nonlocal"\cite{Gol11} abound in scientific publications, and especially in popular literature.  Experts in the field often use the term ``nonlocality" to designate particular non-classical aspects of quantum entanglement and do not confuse the term with superluminal interactions. However, many physicists and philosophers seem to take the term more literally. 

For me the term {\it nonlocality} is so fraught with misinterpretation that I feel we'd all be better off if it were removed from discourse on quantum mechanics. I confess that I'm neither a theoretical physicist nor a philosopher but rather an experimental physicist and observational astronomer and it's from this vantage point that I'll try to convince you that quantum mechanics does not require spooky action at a distance of any kind and will then tell you to what I attribute the magic of quantum mechanical entanglement. But first, a brief review of quantum entanglement and why this phenomenon has led to claims of the nonlocality of quantum mechanics.

\section{Quantum Nonlocality}

Consider two microscopic quantum systems, {\it A} and {\it B}, that interact with each other and then fly off in opposite directions. According to quantum mechanics, the results of measurements made on {\it A} will be correlated with those made on {\it B}. Furthermore, by changing the experimental arrangement used to observe subsystem {\it A}, the correlations with the results of the measurements on remote subsystem {\it B} will also change. The seemingly plausible implication is that the results of the measurements on {\it B} directly depend on the choice of measurements made on {\it A} independent of the distance between the two.\footnote{However, because these correlations can only be interpreted in terms of a quantum mechanical statistical distribution of many such observations, no usable information can be transmitted from {\it A} to {\it B} in this way thereby preserving consistency with special relativity.} This simple argument is specious, as will be clarified shortly. To be sure, Bell's demonstration that any classical, hidden variable treatment of the system demands acausal, superluminal signals provides sufficient grounds to summarily dismiss such classically based models. However, to turn the argument around and assert the nonlocality of nature is not justified. 

So how are the arguments for the nonlocality of quantum mechanics (and nature) constructed? There are a plethora of points of view in the literature.  I offer the following two simply to provide counterpoint to my subsequent account. Physicist/philosopher Abner Shimony puts it like this\cite{Shi13}:
\begin{quote}
Locality is a condition on composite systems with spatially separated constituents, requiring an operator which is the product of operators associated with the individual constituents to be assigned a value which is the product of the values assigned to the factors, and requiring the value assigned to an operator associated with an individual constituent to be independent of what is measured on any other constituent.
\end{quote}
With a little thought, it is evident that Shimony's definition of locality precludes precisely the types of correlations that quantum mechanics demands in which case there is no more to be said. However, I find that this definition of locality to be reminiscent of the classicality that Bell inequality experiments, among many others, have already disproved. In fact, it amounts to little more than defining quantum mechanics to be nonlocal. This is not to say that Shimony does not appreciate the subtleness of designating quantum mechanics to be nonlocal. He continues,
\begin{quote}
Yes, {\it something} is communicated superluminally when measurements are made upon systems characterized by an entangled state, but that something is {\it information}, and there is no Relativistic locality principle which constrains its velocity...This point of view is very successful at accounting for the arbitrarily fast connection between the outcomes of correlated measurements, but it scants the objective features of the quantum state. Especially it scants the fact that {\it the quantum state probabilistically controls the occurrence of actual events.}
\end{quote}
In addition, Shimony acknowledges that nonlocality is far from clear-cut and is closely related to the measurement problem and the concomitant wave function collapse (a topic to which I'll return later).
\begin{quote}
The peculiar kind of causality exhibited when measurements at stations with space-like separation are correlated is a symptom of the slipperiness of the space-time behavior of potentialities. This is the point of view tentatively espoused by the present writer, but admittedly without full understanding. What is crucially missing is a rational account of the relation between potentialities and actualities -- just how the wave function probabilistically controls the occurrence of outcomes. In other words, a real understanding of the position tentatively espoused depends upon a solution to another great problem in the foundations of quantum mechanics -- the problem of reduction of the wave packet.
\end{quote}

Physicist H. W. Wiseman, an expert in quantum information theory, reviewed the history of quantum nonlocality in ``From Einstein's Theorem to Bell's Theorem: A History of Quantum Nonlocality"\cite{Wis06}. Referring to Einstein's autobiographical notes\cite{Ein49}, he concludes that Einstein's logical deduction (from the EPR gedanken experiment) is that one of the following three premises must be false: 1) quantum mechanics is a complete theory; 2) locality as implied by special relativity; or 3) the independent reality of physically separated objects. In their 1935 EPR paper, the conclusion was that quantum mechanics does not provide a complete description of reality. However, Wiseman maintains that after Bell's theorem, the implication is that ``even if statistical QM is not complete...[it] violates locality or reality". It is clear from the first paragraph of Bell's paper\cite{Bel64} that he concurred with this conclusion. The subsequent experimental verification of EPR type entanglement then leaves two possibilities, ``the world is nonlocal -- events happen which violate the principles of relativity or objective reality does not exist -- there is no matter of fact about distant events"\cite{Wis06}. For Wiseman the latter possibility smacks of solipsism. ``Compared to solipsism, the proposition that relativity is not fundamental, and that the world is nonlocal, seems the lesser of two evils."\footnote{Rather than embracing either nonlocality or solipsism, Mermin's\cite{Mer85} more balanced position is simply that the experimental verification of the violation of Bell's inequality provides direct evidence that excludes Einstein's particular concept of an ``independent existence of the physical reality."}  Wiseman concludes that
\begin{quote}
If one assumes the world to be real, then Bell-experiments have proven that it is nonlocal. The nonlocality demonstrated in these experiments does not enable superluminal signalling (and does not allow a preferred reference frame to be identified). It has therefore been called uncontrollable nonlocality, to contrast with (hypothetical) controllable nonlocality which would enable superluminal signaling. But this uncontrollable nonlocality is not purely notional. It can be used to perform tasks which would be impossible in a world conforming to the postulates of relativity. Thus, uncontrollable nonlocality reduces the status of relativity theory from fundamental to phenomenological.
\end{quote}
In a footnote Wiseman adds, ``To be scrupulous, there are perhaps four other ways that the correlations in such an experiment could be explained away." It is the first way that is relevant here because it is related to the position that I put forward below, namely ``One could simply `refuse to consider the correlations mysterious'". Wiseman notes that Bell counters this explanation with\cite{Bel81} ``Outside [the] peculiar context [of quantum philosophy], such an attitude would be dismissed as unscientific. The scientific attitude is that correlations cry out for explanation." I'll return to this sentiment shortly.

Two philosophers who represent the current pro-nonlocality side of the debate are David Albert and Tim Maudlin.
While acknowledging that nonlocal quantum influences ``cannot possibly be exploited to transmit a detectable signal...to carry information, nonlocally, between any two distant points", Albert \cite{Alb92} is unequivocal that ``What Bell has given us is proof that there is as a matter of fact a genuine nonlocality in the actual workings of nature, {\it however} we attempt to describe it, period."  Similarly, Maudlin \cite{Mau11} ``argued that the results are unequivocal:"
\begin{quote}

Violation of Bell's inequality does not require superluminal matter or energy transport.

Violation of Bell's inequality does not entail the possibility of superluminal signaling.

Violation of Bell's inequality does require superluminal causal connections.

Violation of Bell's inequality can be accomplished only if there is superluminal information transmission.
\end{quote}

Taking the nonlocality of quantum mechanics as a given, these two philosophers, in Maudlin's words\cite{Mau11},  ``must turn our attention to the question of the compatability of that non-locality with the relativistic picture of space-time.  We must take up the question of whether a complete quantum theory can be rendered Lorentz invariant."  While their quest has not yet succeeded, the works of these two philosophers critically evaluate several notable attempts.  Two related assumptions that are often made in such deliberations are (even if sometimes unspoken): 1) A fundamental physical theory must provide a complete description of physical reality (Einstein's quest); and 2) The problem of wave function collapse (the measurement problem) needs to be resolved (as Shimony declared).  The present paper challenges the very nonlocality that these authors accept as well as the notions of completeness and wave function collapse (see Section 6).  As a consequence, I will refrain from commenting further on their works.

\section{The Separation Principle}\label{SP}

The conclusion of EPR was not that quantum mechanics is nonlocal nor that objective reality does not exist but rather that quantum mechanics is not a complete description of reality\footnote{ As just mentioned, Bell and Wiseman argue that even if quantum mechanics is incomplete, the quantum mechanical violation of Bell's inequality implies that it either violates the principles of relativity or objective reality does not exist.}. The argument in the EPR paper was based on a demonstration that quantum mechanics cannot simultaneously contain all elements of physical reality. Podolsky wrote the paper and Einstein was less than satisfied with it. Howard\cite{How07} points out that in a letter to Schr\"{o}dinger written a month after the EPR paper was published, Einstein chose to base his argument for incompleteness on what he termed the ``separation principle" and continued to present this argument ``in virtually all subsequent published and
unpublished discussions of the problem."\cite{How07}  According to the separation principle, the real state of affairs in one part of space cannot be affected instantaneously or superluminally by events in a distant part of space. In his letter to Schr\"{o}dinger, Einstein explained\cite{How07}
\begin{quote}
After the collision, the real state of ({\it AB}) consists precisely of the real state {\it A} and the real state of {\it B}, which two states have nothing to do with one another. The real state of {\it B} thus cannot depend upon the kind of measurement I carry out on {\it A} [separation hypothesis]. But then for the same state of {\it B} there are two (in general arbitrarily many) equally justified $\Psi_B$, which contradicts the hypothesis of a one-to-one or complete description of the real states.
\end{quote}

In some ways, I completely agree with a separation principle; however, from my experimentalist perspective, I would interpret it as ``If systems {\it A} and {\it B} are spatially separated, then a measurement of system {\it A} can, in no way, have any effect on any possible measurement of system {\it B}." Whereas Einstein's principle required that a measurement of system {\it A} can have no effect on the {\it state} of system {\it B}, the experimentalist's separation principle requires that a measurement of system {\it A} can have no effect on the {\it result} of any measurement on system {\it B}.  I know of no experimental evidence to the contrary of this principle nor does standard quantum mechanics predict any such violation. It is interesting that in the very first paragraph of his paper Bell defined locality in precisely this way, i.e., ``It is the requirement of locality or more precisely that the result of an experiment on one system be unaffected by operations on a distant system with which it has interacted in the past, that creates the essential difficulty."\cite{Bel64} However, the ``difficulty" only presents itself when applied in conjunction with classical-type hidden variables, which are not present in ordinary quantum mechanics.

As an illustration, consider Bell's original gedanken experiment (due to Bohm and Aharonov\cite{Boh57}): the emission of two oppositely moving spin $\frac{1}{2}$ particles in a singlet state. Their combined wave function is given by
\begin{equation}
	\Psi (1,2)= \frac{1}{\sqrt{2}} \{|1,\uparrow \rangle |2,\downarrow \rangle -  |1,\downarrow \rangle |2,\uparrow \rangle \}_z
\end{equation}
where $\uparrow$ and $\downarrow$ indicate the $z$ components of the spins of particles 1 and 2.  Now suppose that the spin of particle 1 is measured with a Stern-Gerlach apparatus oriented in the $\hat{z}$ direction and is determined to be $\uparrow$. The usual statement is that such a measurement instantaneously collapses the wave function of particle 2 such that $\Psi (2)= |2,\downarrow \rangle_z$. On the other hand, the original wave function can also be expressed as
\begin{equation}
	\Psi (1,2)= \frac{1}{\sqrt{2}} \{|1,\uparrow \rangle |2,\downarrow \rangle -  |1,\downarrow \rangle |2,\uparrow \rangle \}_x
\end{equation}
Then, if the spin of particle 1 is measured with a Stern-Gerlach apparatus oriented in the $\hat{x}$ direction and is determined to be $\uparrow$, the wave function of particle 2 collapses to $\Psi (2)= |2,\downarrow \rangle_x$.  This violates Einstein's separation principle that the measurement of particle 1 can have no effect on the {\it state} of particle 2. Yet, this scenario does not violate the experimentally motivated separation principle. If one measures particle 1 to be $\uparrow$ in any direction, we know the measurement of particle 2 has to be $\downarrow$. This perfect (anti)correlation is built into the two particle system because they are in a singlet state.  Of course one might argue, as did Einstein, that the above scenario provides evidence that the result of the measurement on particle 1 {\it causes} a change in the {\it state} of particle 2.  However, this requires a precise definition of the notion of a {\it state} as well as of {\it cause}.  If one defines {\it state} as the probability distribution of possible outcomes of a measurement of particle 2, then Einstein's separation principle is the same as the experimentalist's separation principle for which there is no conflict with quantum mechanics.  Even if one defines {\it state} in such a way that it is changed in the above scenario, the notion of a {\it cause} offers little in the way of explanation if no physical model of the cause is offered. 

The problem becomes a bit stickier if one measures the spin of particle 1 in an arbitrary direction $\hat{n}$ where $\hat{n} \cdot \hat{z} = \cos  \theta$. It is convenient to express $|1,\uparrow \rangle$ and $ |1,\downarrow \rangle$ in an $\hat{n}$ basis, i.e., 
\begin{equation}
|1,\uparrow \rangle_z	= \cos{\frac{\theta}{2}} |1,\uparrow \rangle_n + \sin{\frac{\theta}{2}} |1,\downarrow \rangle_n
\end{equation}
and
\begin{equation}
|1,\downarrow \rangle_z	= -\sin{\frac{\theta}{2}} |1,\uparrow \rangle_n + \cos{\frac{\theta}{2}} |1,\downarrow \rangle_n
\end{equation}
It is straightforward to show that the correlation of the measured components of the spins of the two particles in the $\hat{n}$ and $\hat{z}$ directions is given by $-\cos{\theta}$ (with the spins in units of $\hbar / 2)$. That is, the correlations between the measurements of the two distant systems are changed, which may seem to be a problem for any separation principle. How is it that the original states of the two particles know about the perfect anti-correlation, this new correlation and, in fact, about every correlation of all conceivable measurements made on the two particles? Well, that's the magic of quantum mechanics but it does not violate the experimentalist separation principle.

Suppose observer 1 (the observer of particle 1) chooses an apparatus to measure the spin in the $\hat{n}$ direction and that choice were somehow immediately (superluminally) transmitted to and affects the measurement of particle 2 in the $\hat{z}$ direction a moment latter. If the events that define these two measurements have a space-like separation, as in the EPR experiment, it is not possible for them to be unambiguously time ordered. Because of their space-like separation, it is straightforward to identify another inertial observer who will claim that the observation of the spin of particle 2 occurs well before that of particle 1. Then either knowledge of the choice of the experimental arrangement of observer 1 must be communicated backward in time to observer 2, or the results of the measurement of particle 2 must be subsequently communicated to 1 and determine observer 1's choice of the experimental arrangement. The former certainly wreaks havoc with causality while the latter seriously compromises free will. In fact, one needn't appeal to other inertial observers. Within a common rest frame of the two observations, simply have the observation of 2 take place well before (i.e., much closer to the interaction region) than the observation of 1. Now suppose that the result of that measurement is sent via a light signal to a third party located midway between the two observers and is recorded in a lab notebook well before observer 1 makes the choice of the direction $\hat{n}$. Then it is absolute clear that that choice of measurement of 1 could have no possible effect on the measurement of 2, superluminal or otherwise, as is consistent with the experimentalist separation principle. After all, the measurement of 2 is ``written in stone" before the choice of $\hat{n}$ is made. The correlations of many such measurements will be those predicted by quantum mechanics regardless of the choice of measurements. All those correlations are built into the entangled wave function of the two particles. As for the measurements of particle 1 and 2 themselves, they have absolutely no effect on each other. There is no spooky action at a distance.

\section{Correlations in Single Particle States}\label{prob}

Okay, so how do I explain the magic of the correlations of entangled quantum mechanical systems?  First, a quick review of Bell's theorem.  Let $P(\hat{a},\hat{b})$ denote the correlation of spin measurements of particles 1 and 2 along the directions $\hat{a}$ and $\hat{b}$ respectively (Bell's notation).  Then the inequality derived by Bell (roughly a one page proof), $1+P(\hat{b},\hat{c}) \geq |P(\hat{a},\hat{b})-P(\hat{a},\hat{c})|$, is the condition that must be satisfied by any classical, local, hidden variable description of the hypothetical singlet spin system.  Suppose that $\hat{b}$ is in the $\hat{z}$ direction while $\hat{a}$ and $\hat{c}$ are in directions that are $\pm \theta$ from the $\hat{z}$ direction.  Then from the quantum mechanical correlation analysis above, Bell's inequality takes the form $1-\cos{\theta}\ge |\cos 2\theta - \cos \theta |$, which is clearly violated for $0 < \theta < \pi/2$ (the domain of applicability of the inequality).  The implication is that quantum mechanics is inconsistent with any classical, local, hidden variable theory.

The violation of the Bell inequality arises from the quantum mechanical prediction of the correlations of measurements of the spins, i.e., $P(\hat{z},\hat{n})=-\cos \theta$, but this correlation arises directly from the single particle wave function.  Suppose a spin $\frac{1}{2}$ particle is in the $|\uparrow\rangle_z$ state.  This can be written in the $\hat{n}$ basis as $|1,\uparrow \rangle_z = \cos{\frac{\theta}{2}} |1,\uparrow \rangle_n + \sin{\frac{\theta}{2}} |1,\downarrow \rangle_n$.  Now consider the correlation between a measurement of the particle's spin in the $\hat{z}$ direction and a hypothetical measurement of the spin of the same particle in the $\hat{n}$ direction.  Then
\begin{equation}
P(\hat{z},\hat{n}) = \langle \uparrow|_z \sigma_z \sigma_n (\cos{\frac{\theta}{2}} |\uparrow \rangle_n + \sin{\frac{\theta}{2}}|\downarrow \rangle_n)=\cos\theta
\end{equation}
where $\sigma_{z}$ and $\sigma_{n}$ are the spin operators in the $\hat{z}$ and  $\hat{n}$ directions. 
This is the same correlation as for Bell's gedanken experiment except for the minus sign, which is due to the singlet state of Bell's two particles.  One can easily make them exactly the same simply by requiring either of the measurements to change the sign of the output.  So it would {\it seem} (more on this later) that this hypothetical measurement also violates Bell's inequality even though there is no question of nonlocality, superluminal interactions, or even entanglement.  After all, there's only a single particle.  Of course, the measurement of the spin in the $\hat{n}$ direction is only hypothetical because one cannot simultaneously measure components corresponding to non-commuting operators.  Philosophers refer to the results of such a measurement as {\it counterfactually definite}, a notion that is uncontroversial in deterministic classical physics.  Bell's theorem clearly makes use of counterfactual definiteness; his inequality involves the correlations of the spins of the two particles in each of two different directions that correspond to non-commuting spin components. This use of counterfactuals is entirely appropriate because it is used to investigate a test for classical hidden variable theories.    Even so, I realize that one might object to such a counterfactual gedanken experiment in the present single particle scenario. However, the purpose of this hypothetical experiment is to demonstrate that the resulting correlation has little to do with entanglement, nonlocality, and superluminal signals and everything to do with superposition and the nature of quantum states, whether entangled or not.  One can render the counterfactual measurement to be an actual measurement by cloning the original single particle state and performing an actual measurement on the cloned state.  In effect, this is precisely what Bell's singlet entangled state accomplishes.

In fact, one needn't couch the discussion in terms of a counterfactual measurement.  Suppose we simply prepare an isolated system in a $|\uparrow\rangle_z$ state and then subsequently measure the spin in the $\hat{n}$ direction.  The mean value of the measured spin would again be the same $\cos{\theta}$ as in Bell's gedanken experiment.  This follows directly from the single particle decomposition in Eq. 3.3.  Bell's spin singlet system might well be considered to be a prescription for preparing such a single particle state but this in no way indicates that the remote measurement of particle 2 has any effect whatsoever on the measurement of particle 1.  The latter follows directly from the quantum properties of the single particle wave function.

Yet another way of posing the single particle scenario is to consider two actual measurements of the spin in the $\hat{z}$ and $\hat{n}$ directions.  The experimental arrangement is a Stern-Gerlach apparatus aligned in the $\hat{z}$ direction with another Stern-Gerlach apparatus located in the upper arm of the former but aligned in the $\hat{n}$ direction (see Fig. 1). 
\begin{figure}[htb]
\vbox{\hfil\scalebox{1.0}
{\includegraphics{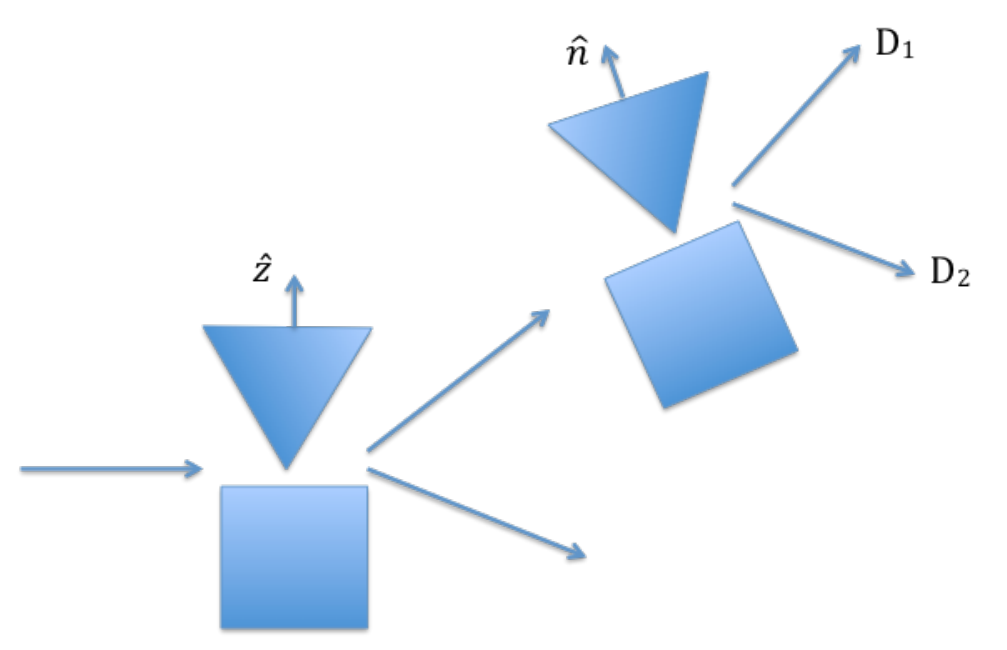}}\hfil}
{\caption{\footnotesize{Double Stern-Gerlach Apparatus}}}
\end{figure}
Electron counters, D1 and D2, are placed at the two outputs of the 2nd apparatus.    If there is a detection in either of these two outputs, then we know that the state of the electron incident on the 2nd apparatus is $|\uparrow \rangle_z$ and so can consider this to be a measurement of the $\hat{z}$ component of the spin of the incident electron.  On the other hand, $|\uparrow \rangle_z$ can be represented as $|1,\uparrow \rangle_z = \cos{\frac{\theta}{2}} |1,\uparrow \rangle_n + \sin{\frac{\theta}{2}} |1,\downarrow \rangle_n$ and so the specific counter that registers the detection can be considered a measurement of the $\hat{n}$ component of the spin.  Then the correlation of the spin of the incident electron, +1, with the spin detected by the second apparatus is again given by $P(\hat{z},\hat{n})=\cos \theta$.  In this case the final step of the two measurements performed on a single particle occur simultaneously and so nonlocality, superluminal wave function collapse, and entanglement are all irrelevant.

While the single particle scenario results in the same correlations as in the two particle spin singlet system of Bell, does this imply that the former also constitutes a violation of a Bell-type inequality and thereby demonstrates the fallacy of hidden variable theories?  The answer is, in general, no.  Bell \cite{Bel66} pointed out that the single particle case does provides an example of von Neumann's ``no hidden variables" proof; however, that proof involved an assumption that Bell and others (e.g., Mermin\cite{Mer93}) characterized as ``silly".  Bell went on to construct a viable hidden variable theory for a single spin $\frac{1}{2}$ particle that didn't rely on von Neumann's silly condition.\footnote{ Mermin \cite{Mer93} provided a simpler version of this construction.}  Therefore, the single particle case is not particularly useful in evaluating the viability of hidden variable theories nor illuminating arguments for the nonlocality of quantum mechanics.  On the other hand, the two measurement Stern-Gerlach scenario above can be couched in terms of the four-dimensional , two particle subject of the Bell-Kochen-Specker theorem\cite{Mer93} that illustrates the incompatibility of quantum mechanics with hidden variable theories contingent on the assumption of ``noncontextuality"\footnote{Noncontextuality is the assumption that a hidden variable theory must assign to an observable a value that is independent of the complete disposition of the relevant measuring apparatus.} of the latter.  Bell\cite{Bel66} also considered this assumption unreasonable but Mermin would not go so far \cite{Mer93}.  While one might argue with applying the Bell-Kochen-Specker theorem to the above two measurement single particle scenario, I remind the reader that the purpose of the single particle scenarios is not to comment on the viability of hidden variable theories {\it per se} but rather to shed light on the origin of the quantum mechanical correlations that figured so critically in Bell's  proof and thereby to help explain the magic of quantum mechanical entanglement.  The point is that the two particles in Bell's spin singlet system each carry with them the single particle correlations that are eventually manifested in correlated observations of the entangled state. That is, the magic of 
the correlations of Bell's entangled system is a direct consequence of the quantum behavior of a single spin $\frac{1}{2}$ particle.

\section{Entanglement and Quantum Nonlocality}\label{EQN}

As I pointed out above, Bell's definition of locality is essentially the same as the experimentally motived separation principle, which I claim is not violated by quantum mechanics.  So why did he find that quantum mechanics violates his inequality and, hence, the notion of locality?  The problem comes from the translation of his verbal definition of locality into its mathematical expression in the context of hidden variables.  In his notation, let $A(\hat{a},\lambda)$ and $B(\hat{b},\lambda)$ represent the results of the spin measurements of particles 1 and 2 in the $\hat{a}$ and $\hat{b}$ directions respectively.  Then Bell expresses the expectation value of the product of (the correlation of) the two measurements as $P(\hat{a},\hat{b}) = \int d\lambda \rho(\lambda)A(\hat{a},\lambda)B(\hat{b},\lambda)$ where $\lambda$ represents any hidden variables with statistical distribution $\rho(\lambda)$.  This is analogous to the product of operators definition (given by Shimony) that precludes the types of correlations that quantum mechanics demands.  Therefore, Bell's mathematical expression of locality goes beyond his initially stated experimentalist separation principle.

To be sure, in his 1964 paper, Bell did not conclude that quantum mechanics is nonlocal, only that a classical hidden variable model designed to reproduce the statistical predictions of quantum mechanics must necessarily be nonlocal.  However, in a subsequent paper\cite{Bel75}, he formalized ``a notion of local causality" that was directly applicable to quantum mechanics and concluded that quantum mechanics itself is nonlocal.  This analysis used a mathematical expression of locality similar to the above direct product of probabilities. The other concept necessary for the derivation of Bell's original inequality was a notion of ``classical realism", which in the 1964 paper takes the form of classical hidden variables and the counterfactual definiteness they imply.\footnote{The standard Copenhagen interpretation of quantum mechanics is counterfactually indefinite because one cannot make definite statements about the hypothetical results of measurements of quantities corresponding to non-commuting operators.} In his 1975 paper he introduced the more general concept of ``local beables".  However, some of these beables functioned in the same way as the hidden variables in his 1964 paper and so it is not surprising that an inequality he derived from these assumptions is violated by the predictions of quantum mechanics thereby revealing that quantum mechanics violated locality as he defined it.

The phenomenon of entanglement is not restricted to quantum mechanics.  Two classical systems that interact with each other before moving off in different directions are also entangled.  To the extent that the interaction can be completely characterized, one can predict the correlations of all possible measurements made on the two systems whether space-like separated or not. Furry discussed a classical entangled system involving cards in boxes and concluded\cite{Fur62}:
\begin{quote}
There is no contradiction with relativity, and the attaining of information from one place or the other is just what it sounds like. The difference, of course, between the classical and the quantum picture is that the quantum mechanical state does not correspond to this because this nice classical picture of the box with two envelopes is the hidden parameter description and the hidden parameter description is denied in quantum mechanics. But this is the only difference between the two things and there is no difference at all about the questions of information and of distance and time.
\end{quote}
What distinguishes quantum entanglement from its classical counterpart is simply the superposition of states and the quantum interference it implies.

There is certainly no doubt that quantum entanglement is a much richer phenomenon than its classical counterpart, cf. quantum information and quantum computing.  However, we all know that quantum mechanics is, in general, much richer than classical mechanics.  Quantum theory is capable of describing atoms and their interactions, the properties of solids, liquids, and gases, and is applicable not only to physical systems but to chemistry and, by inference, biology as well, whereas classical mechanics has much less to offer for such systems. Quantum entanglement, as Mermin declared\cite{Mer85}, ``...is as close to magic as any physical phenomenon...and magic should be enjoyed."  What I have endeavored to demonstrate is that Bell's theorem and what it reveals have little to do with quantum nonlocality, superluminal propagation, and even entanglement.  Rather, they follow directly from the single particle quantum mechanical superposition, $|1,\uparrow \rangle_z = \cos{\frac{\theta}{2}} |1,\uparrow \rangle_n + \sin{\frac{\theta}{2}} |1,\downarrow \rangle_n$.  The simple examples above are intended to demonstrate that the correlations of entangled states can be understood in terms of the standard quantum superposition of any system whether entangled or not.

So I suppose that my account is what Wiseman called a refusal to consider the correlations mysterious.  However, that's not quite right.  I absolutely think the correlations are mysterious, that is, I don't have ``rocks in my head".  I simply embrace the mysteries of quantum mechanics.  Rather, it's the nonlocality view that endeavors to force a kind of classicism on phenomena that are patently non-classical.  As for Bell's exhortation that ``the scientific attitude is that correlations cry out for explanation", I absolutely agree.  However, as I frequently have to remind my students, {\it correlation} does not necessarily mean {\it causation}.  Rather my explanation for the observed correlations of entangled states is the quantum behavior of matter and radiation.

\section{The Incompleteness of Quantum Mechanics} \label{IQM}

The conclusion of the EPR paper was neither that quantum mechanics yields incorrect predictions nor that quantum mechanics is nonlocal.  Rather it was that quantum mechanics does not provide a complete description of reality. The argument in EPR, while peripherally invoking the reduction of the wave function, did not even mention the concept of locality but rather involved a rather byzantine analysis involving a definition of elements of reality. The EPR manuscript was written in Einstein's absence by Podolsky and Einstein was not happy with it.  In particular, it had ``not come out as well as I really wanted; on the contrary, the main point was, so to speak, buried by the erudition"\cite{How07}.  Rather, Einstein based his argument for incompleteness on his principle of separation (as discussed above), which explicitly invokes both locality and the reduction of the wave function.

For the reasons stated above, I don't find this argument convincing; however, Einstein was certainly not alone in his contention that quantum mechanics does not provide a complete description of reality.  Many prominent physicists have spoken to the incompleteness of quantum mechanics including Pauli, Dyson, Furry, Wigner, and others.  In a sense even Bohr, Einstein's chief adversary in these matters, acknowledged the incompleteness of quantum mechanics by insisting that measurements must necessarily be described in ordinary language outside the formalism of quantum mechanics.  Dyson\cite{Dys02} made the incompleteness more explicit through four sensible gedanken experiments, couched in the ordinary language of measurements and experimental physics, that defy quantum mechanical explanation.  Quantum mechanics quite simply cannot be applied to all conceivable situations.  For Dyson, the dividing line between classical and quantum physics is the same as the dividing line between the past and the future.  The wave function constitutes a statistical prediction of future events.  After the event occurs, the wave function doesn't collapse rather it becomes irrelevant.\footnote{Wigner\cite{Wig62} made essentially the same point.}   The results of past observations, facts, are the domain of classical physics; the statistical probabilities of future events, wave functions, are the domain of quantum physics.  Ergo, quantum mechanics does not provide a complete description of nature.

It's interesting to consider Bell's theorem in this context.  Dyson demonstrated that the inappropriate application of quantum mechanics to the motion of a particle in the past results in a violation of the Heisenberg uncertainty principle.  The flip side is that the inappropriate application of a classical notion of locality results in a violation of Bell's inequality for future events and the concomitant claim of quantum nonlocality\cite{Dys16}.

Wigner\cite{Wig62} and Furry\cite{Fur62} made the Bohrian argument that the notions of measurement and experiment necessarily fall outside the realm of quantum mechanics and for that matter outside the realm of classical mechanics.  Stapp chose to emphasize this pragmatic view of experiments by using the word specifications, i.e.,\cite{Sta72}
\begin{quotation}
Specifications are what architects and builders, and mechanics and machinists, use to communicate to one another conditions on the concrete social realities or actualities that bind their lives together. It is hard to think of a theoretical concept that could have a more objective meaning. Specifications are described in technical jargon that is an extension of everyday language. This language may incorporate concepts from classical physics. But this fact in no way implies that these concepts are valid beyond the realm in which they are used by technicians.
\end{quotation}
Quantum mechanics pays scant attention to measurement other than the (not universal) contention that measurements cause the wave function to collapse.  Even the EPR paper invokes the notion of wave function collapse.  The problem is that wave function collapse, even if (contrary to Dyson's and Wigner's contentions) it were a sensible concept, is not predicted by quantum mechanics. It is an event that happens outside the mathematical formalism of the theory. In fact, formal quantum theory says absolutely nothing at all about measurements and how they should be performed or even how to interpret the results of measurements. The statistical distributions of quantum mechanical predictions follow from rules for how to interpret the constructs of quantum theory in terms of the results of experiments. The measurements themselves are not described by quantum theory but rather by the operational prescriptions of the experimentalists and technicians who perform them. This is yet another sense in which quantum mechanics is incomplete.

Finally, Einstein was willing to consider another resolution to the EPR ``paradox" if one were to assume that quantum mechanics does not apply to a single system but rather only to an ensemble of similarly prepared systems.  In his words, ``The $\Psi$ function does not in any way describe a condition which could be that of a single system; it relates rather to many systems, to `an ensemble of systems' in the sense of statistical mechanics."\cite{Ein36}  The ensemble interpretation, while minimalist, is considered by many to be the most reasonable interpretation of quantum mechanics (e.g., see Ballentine\cite{Bal70}).  While Einstein admitted that ``such an interpretation eliminates also the [EPR] paradox", he added ``To believe this is logically possible without contradiction; but, it is so very contrary to my scientific instinct that I cannot forego the search for a more complete conception." In other words, he would still strongly suspect that quantum mechanics is incomplete.

\section{Final Remarks} \label{FR}

In this paper I argue that the ``magic" of entangled quantum states has little to do with nonlocality, superluminal interactions, or even entanglement itself but rather is yet another example of quantum mechanical superposition, which is magic even in the context of a single particle.  On the other hand, I have no real criticism of Bell's 1964 proof.  So what is it that Bell's theorem is telling us?  For me, it is a nifty demonstration of why it is that hidden variable theories of quantum mechanics do not work. As such it can serve, if properly presented, as a useful pedagogical tool to help beginning students of quantum mechanics overcome an initial tendency to view the ``wave function" as simply a replacement for the notion of ``particle" in the description of nature.  However, for me Furry's 1936 response to the EPR paper and to Bohr's published reply provides a more straightforward demonstration, in the context of two interacting (entangled) systems, of the difference between an intuitive classical minded model that students often conjure and what quantum mechanics actually says.  Furry wrote his paper only a few months after the EPR-Bohr exchange and Bell cited it in the very first sentence of his 1964 paper.

Furry\cite{Fur36a} considers the wave function for two systems ($I$ and $II$) that interacted in the past and now have ceased to interact.  ``One can show that there always exists an expansion, which is in general unique, in the form
\begin{equation}
\Psi(x_1,x_2) = \sum_k {w_k}^{\frac{1}{2}} \phi_k (x_1) \xi_k (x_2)
\end{equation}
where the $\phi_k$ are eigenfunctions of an observable {\it L}... and the $\xi_k$ are eigenfunctions of an observable {\it R}..."  Furry's classically minded hypothesis is that ``during the interaction of the two systems each system made a transition to a definite state, in which it now is, system $I$ being in one of the states $\phi_k$ and system $II$ in one of the states $\xi_k$."  There is no way to predict into which pair of states, $\phi_k\xi_k$ , the systems end up, only that the probability of ending up in any of the pairs is given by $w_k$.  This sort of picture often finds itself in the minds of beginning students of quantum mechanics.  It is one that preserves the primary status of the wave function and associated probabilistic nature of quantum mechanics and yet allows the classical notion that spatially separated systems have independent properties even if these properties are to be specified entirely by quantum mechanical wave functions.  As such, it is a reasonable model for the type of reality, i.e., the real properties of systems, that Einstein wanted to preserve.  Furry then proceeds to demonstrate that while this hypothesis gives the same answers as quantum mechanics for most of the questions raised in discussions of the theory of measurements, there is a general class of questions where contradictions occur, namely those measurements involving ``the well-known phenomenon of `interference' between probability amplitudes."  Furry concluded that ``...there is of course nothing 'magical' about the affair. The interference effect does not come in unless there has been an actual opportunity for the two particles to get interchanged, just as in the case in hand there is never any relation between the systems without the existence of an actual dynamical interaction to start it."\cite{Fur36b}  Furry's classically motivated hypothesis results in the same product of probabilities that Bell and Shimony use as a condition of locality and so it is no mystery why Furry's hypothesis results in the same disagreement with standard quantum interference.

Furry's analysis is in some ways more general than Bell's in that he doesn't specify a particular type of quantum state and in some ways less general in that he specifies a particular classically minded hypothesis rather than some general specification of hidden variables.  For me, Furry's analysis is the better pedagogical tool for helping students of quantum mechanics realize the extent to which quantum mechanics diverges from our classical intuition.  If so, then why is Bell's paper so well known while Furry's paper is largely forgotten?\footnote{Bell's 1964 paper has been cited more than 6000 times in the last 10 years compared to only a handful of citations of Furry's 1936 paper in that time period.}   Perhaps one reason is that Bell derived a specific test, in the form of an inequality, that local hidden variable theories satisfy and that quantum mechanics does not.   Another is that by 1964, Bell was able to conceive of actual experiments that could perform this test, the first of which happened 8 years later.\footnote{Also, these experiments proved to be seminal in the emerging field of quantum information and quantum computing.}   Finally, I'm sure that part of the reason is the subsequent analyses by Bell and others that supposedly demonstrated the nonlocal character of quantum mechanics and by inference of nature herself.  It is with this conclusion that I take issue.
    
In the first paragraph of the Introduction I noted David Mermin's contention that physicists' views of Bell's theorem fall ``between the majority position of near indifference and the minority position of wild extravagance"\cite{Mer85} and it is perhaps not surprising that most papers on Bell's theorem tended toward the minority position and the implied nonlocality of quantum mechanics.  Even so, the contrarian view I have taken is certainly not solitary.  Most physicists seem to prefer to remain silent (the near indifferent majority); however,  some notable scientists, including Gell-Mann, Dyson, Mermin, Furry, Hartle and Peres, while emphasizing different aspects of entanglement, have expressed sentiments similar to mine.  There are also several philosophers among the skeptics.  In his paper on the epistemological implications of Bell's Inequality\cite{Van82}, Van Fraassen points out that most quantum nonlocality arguments hinge on applying the notion of ``common cause" to Bell-type scenarios.  He concludes that
\begin{quote}
...empirical adequacy of a theory consists in it having a model that {\it all} the (models of) {\it actual} phenomena will fit into.  In some cases, the methodological tactic of developing a causal theory will achieve this aim of empirical adequacy, in other case it will not, and that is just the way the world is.  The causal terminology is descriptive, in any case, not of the (models of the) phenomena, but of the proffered theoretical models.  So pervasive has been the success of causal models in the past, especially in a rather schematic way at a folk-scientific level, that a mythical picture of causal processes got a grip on our imagination.
\end{quote}
Arthur Fine\cite{Fin89} points out that the notion of nonlocality arises in the context of seeking an explanation for the correlations of entangled state but then questions whether such correlations even need to be explained.  He argues convincingly ``...that it is only the combination of strong locality with determinism from which the satisfaction of the Bell inequalities follows."  On the other hand, quantum mechanics is decidedly indeterministic and so there is no path from the violation of Bell's inequalities to the nonlocality of quantum mechanics.  One might worry that this indeterminism would be a problem for the strict probabilistic laws that quantum mechanics predicts for repeated measurements on an ensemble of similarly prepared, individual systems.  We have apparently become comfortable with this situation so ``Why, from, an indeterminist perspective should the fact that there is a [correlated] pattern {\it between} random sequences require any more explaining than the fact that there is a pattern internal to the sequences themselves?"  Fine answers this question as follows,
\begin{quote}
The search for ``influences" or for common causes is an enterprise external to quantum theory.  It is a project that stands on the outside and asks whether we can supplement the theory in such a way as to satisfy certain {\it a priori} demands on explanatory adequacy.  Among these demands is that stable correlations require explaining, that there must be some detailed account of how they are built up, or sustained, over time and space.  In the face of this demand, the correlations of the quantum theory can seem anomalous, even mysterious.  But this demand represents an explanatory ideal rooted outside the quantum theory, one learned and taught in the context of a different kind of physical thinking...The quantum theory takes for granted not only that sequences of individually undetermined events may show strict overall patterns, it also takes for granted that such patterns may arise between the matched events in two such sequences.  From the perspective of quantum theory, this is neither surprising nor puzzling.  It is the normal and ordinary state of affairs.
\end{quote}
So Fine might characterize Bell's previously quoted exhortation that ``correlations cry out for explanation" as one that was learned in the context of a different kind of (classical) physical thinking and therefore not appropriate in a discussion of quantum entanglement.

I find the arguments of these authors convincing as is undoubtedly evident from the present paper.  My arguments have been less general and more pragmatic than their more theoretical and philosophical treatments.  I trust that this experimentalist's perspective will make a useful contribution to the dialog about Bell's theorem and quantum nonlocality.

\end{document}